\documentclass[12pt,oneside,reqno]{amsart}

\usepackage[dvipdfmx]{hyperref}
\usepackage{geometry}             
\usepackage{graphicx}
\usepackage{amsmath, amssymb, amsthm, mathrsfs, mathtools, amscd}
\usepackage{color,bm,bbm}
\usepackage{url}
\usepackage{doi}

\usepackage{enumerate}
\usepackage{times}
\usepackage{cite}
\usepackage[normalem]{ulem}

\newcommand{\deq}[1]{ \begin{align}#1\end{align}}
\newcommand{\deqs}[1]{ \begin{align*}#1\end{align*}}
\newtheorem{Theorem}{Theorem}[section]
\newtheorem{Corollary}[Theorem]{Corollary}
\newcommand{\bTheorem}{\begin{Theorem}}
\newcommand{\eTheorem}{\end{Theorem}}
\newcommand{\bCorollary}{\begin{Corollary}}
\newcommand{\eCorollary}{\end{Corollary}}
\newcommand{\bProof }{\begin{proof}}
\newcommand{\eProof}{\end{proof}}
\newcommand{\al}{\alpha}
\newcommand{\id}{{\rm id}}
\newcommand{\Eq}[1]{Eq.~(\ref{#1})}
\newcommand{\If}{\Leftarrow}
\newcommand{\THEN}{\Rightarrow}
\newcommand{\Then}{\Rightarrow}
\newcommand{\Tr}{{\rm Tr}}
\newcommand{\eq}[1]{(\ref{#1})}
\newcommand{\bra}[1]{\langle#1|}
\newcommand{\ket}[1]{|#1\rangle}
\newcommand{\ketbra}[1]{\ket{#1}\bra{#1}}
\newcommand{\cA}{{\mathcal A}}
\newcommand{\cC}{{\mathcal C}}
\newcommand{\cF}{{\mathcal F}}
\newcommand{\cH}{{\mathcal H}}
\newcommand{\cI}{{\mathcal I}}
\newcommand{\cP}{{\mathcal P}}
\newcommand{\cR}{{\mathcal R}}
\renewcommand{\And}{\wedge}
\newcommand{\Or}{\vee}
\newcommand{\Not}[1]{#1^{\perp}}
\newcommand{\Inf}{\bigwedge}
\newcommand{\Sup}{\bigvee}
\newcommand{\p}{{}^{\perp}}
\newcommand{\com}{{\rm com}}

\title[Human logic and cognitive psychology]{
Nondistributivity of human logic and violation of response replicability effect in cognitive psychology}
\author[M. Ozawa and A. Khrennikov]{Masanao Ozawa\\
Center for Mathematical Science and Artificial Intelligence, Academy of Emerging Sciences, 
Chubu University, 1200 Matsumoto-cho, Kasugai 487-8501, Japan\\
Graduate School of Informatics, Nagoya University, Chikusa-ku, Nagoya 464-8601, Japan\\
Andrei Khrennikov\\ 
Linnaeus University, International Center for Mathematical Modeling\\  in Physics and Cognitive Sciences
 V\"axj\"o, SE-351 95, Sweden}
\date{}                     

\begin{document}
\begin{abstract} The aim of this paper is to promote quantum logic as one of the basic tools for analyzing human reasoning. We compare it with classical (Boolean) logic and highlight the role of  violation of the distributive law for conjunction and disjunction. It is well known that nondistributivity is equivalent to incompatibility of logical variables -- the impossibility to assign jointly the two-valued truth values to these variables.  A natural question arises as to whether quantum logical nondistributivity in human logic can be tested experimentally. We show that testing the response replicability effect (RRE) in cognitive psychology is equivalent to testing nondistributivity --under the prevailing conjecture that the mental state update generated by observation is described as orthogonal projection of the mental state vector (the projective update conjecture of Wang and Busemeyer). A simple test of RRE is suggested. In contrast to the previous works in quantum-like modeling, we proceed in the state-dependent framework; in particular, distributivity, compatibility, and RRE are considered in a fixed mental state.  In this framework,  we improve the previous result on the impossibility to combine question order and response replicability effects by using  (von Neumann-L\"uders) projective measurements.
\medskip

\end{abstract}
\maketitle
\thispagestyle{myheadings}
\tableofcontents

\section{Introduction}

In this paper, we promote quantum logic as one of the basic tools for analyzing human reasoning. We emphasize its distinguishing features, especially nondistributivity and incompatibility for some logical variables. Quantum reasoning can be associated with  heuristic and emotional fast thinking considered by  Kahneman \cite{KAH}, though this is merely an association at present.   

We recall that, since the work of Boole \cite{Boole}, classical Boolean logic has been used as the basic logical tool for human reasoning. Nowadays, Boolean logic is also important for
symbolic AI which is based on propositional representations of knowledge.
This is a logic represented by the algebra of propositions endowed with the operations of conjunction, disjunction, and negation. For our further consideration, it is important to stress that the first two operations satisfy the distributive law. {\it Boolean logic is distributive.} The basic mathematical representation of  Boolean algebra is the set algebra with the operations of intersection, union, and complement. 

Quantum logic, originally introduced by Birkhoff and von Neumann \cite{BvN36}, is, on the other hand,  mathematically represented as the lattice of projections on a complex Hilbert space;
equivalently, it is represented as the lattice of closed subspaces in a complex Hilbert space,
and those two lattices are isomorphic by the correspondence between a subspace and the projection onto that subspace.
Quantum logic is endowed with the corresponding logical operations (conjunction, disjunction, and negation). 
But, interrelation between the logical operations differs from the classical 
case. The main difference is in violation of  the distributive law. Quantum logic relaxes this important law of  classical logic to the orthomodular law \cite{Kal83}
and this leads to more general rules of reasoning. (See, e.g., \cite{Aumann} on violation of Aumann's theorem about impossibility to agree or disagree by quantum decision makers.) 
As is known (Theorem A), nondistributivity of operating with propositions is equivalent to their incompatibility -- the impossibility of assigning the two-valued truth values to logical variables -- even in the state dependent formulation. 
Can we check the quantumness of human logic  explicitly? 
The natural question arises:     

{\it Can the distributive law in human reasoning be checked experimentally?} 

Naturally, this sounds a difficult question to answer, because logical laws are used in a very deep level of human information processing even unconsciously.
We show that the {\em response replicability effect} (RRE) known in cognitive psychology might be useful to find the answer to these questions.

We recall that recently the problem of mathematical modeling of psychological effects was intensively studied in the framework of quantum measurement theory 
\cite{VN,DL70,Dav76,Oza84,Oza97,Oza04}
(for general theory see, e.g., an introductory survey \cite{Oza21} and references therein; for its application in cognition, psychology, social and political sciences see, e.g., monographs \cite{QL0zz,QL1,BB12,BAF,QL4zz,QL5zz}, recent reviews \cite{BPO,BKO} and references therein; we also mention a few recent articles \cite{BDZ}-\cite{FX}). 
In particular, the problem of the possibility to combine of the {\it question order effect} (QOE)
and the {\it response replicability effect} (RRE)  attracted a lot of interest 
\cite{PLOS,FOUND,ENTROPY,MATHPSYCH}. 

Here, QOE  is an effect of the  dependence of the sequential joint probability distribution of answers on the questions' order: $p_{AB}\not=p_{BA}$.
On the other hand, RRE concerns correlations for the answers to sequential questions.
Suppose that after answering the $A$-question with the ``yes'', Alice is asked another question $B$, and gives an answer to it. And then she is asked $A$ again. In the social opinion pools and other natural decision making experiments, Alice would definitely repeat her original answer to $A$, ``yes''. This is $A-B-A$ response replicability. (In the absence of $B$-question, we get $A-A$ replicability). The combination of $A-B-A$ and $B-A-B$ replicability forms RRE.

QOE has been intensively studied in psychology (see, e.g.,  \cite{Moo02}, where the typical opinion poll data sets that show QOE are  discussed and classified; see also the references therein). 
Its modeling on the basis of the quantum  formalism \cite{WB13,WSSB14,PO0} stimulated development of the quantum cognition project.
In particular, a pioneering work by Wang and Busemeyer \cite{WB13} models two questions by two noncommuting projections.
This modeling naturally leads to the assumption that the logical structure of human reasoning about those two questions is based on non-distributive quantum logic. 

Subsequently, the role of RRE was highlighted in this project \cite{PLOS,FOUND,ENTROPY,MATHPSYCH};  before
psychologists did not investigate RRE and its role in functioning of human cognition. 
RRE is generally considered to hold in the poll, if the respondent
is rational and has an adequate memory.
In \cite{PLOS}, RRE was coupled to QOE as the obstacle to describing QOE with the projection-type (von Neumann-L\"uders) quantum instruments. We recall that a quantum instrument \cite{DL70,Dav76,Oza84,Oza97,Oza04} is a mathematical
representative for a general quantum measurement that describes both the probability of the measurement
outcome and the state update given by any possible outcomes. 
The latter plays the crucial role in the problem of combination of QOE and RRE \cite{ENTROPY,MATHPSYCH}. 

It seems that RRE has not been subject to much empirical evaluation. To the
best of our knowledge, a single experiment  \cite{BWRRE} has been performed to test the conclusion put forward in \cite{PLOS}.
The interpretation of  experimental statistical data by the authors of article \cite{BWRRE} was questioned in a few comments to this paper (cf.~online comments to \cite{BWRRE}). 

In this paper, we highlight RRE by coupling it with logic of human reasoning. 
Under the prevailing conjecture (see Wang and Busemeyer \cite{WB13}) that brain's self-observations generate the mental state update of the projection
type (the projective update conjecture), RRE implies distributivity, which contradicts QOE. We note that under the projective update conjecture, RRE and QOE are mutually exclusive. Thus, the testable assumption RRE+QOE negates experimentally the projective update conjecture.

Under the projective update conjecture, Theorem A implies that the above question on distributivity law of human logic is coupled to another important question:

{\it How can it be shown that mental observables (say questions) are incompatible?} 
 
In physics, the question of incompatibility of the basic quantum observables, say position and momentum, can be easily solved 
in the theoretical framework. Here quantum observables are generated from classical phase space mechanics, the Hamilton mechanics, via the quantization  procedure,
in which the quantum commutator corresponds to the classical Poisson bracket. 
For cognition, we cannot proceed in this way, since  the mental analog of 
phase space mechanics has not been created.

Now we point to another  novel invention in this paper. In physics, the use of space geometry gives the possibility to generate 
a large variety of  quantum states, e.g., by using polarization beam splitters oriented in different directions. In psychology and 
cognition, the class of mental states which can be prepared for measurement of concrete mental observables is very restricted.
Here it is more natural to study {\it state dependent features} of observables (see \cite{Oza04,05PCN,06QPC,16A2,19A1} for such an approach in physics).  In this paper, we introduce the notions of state dependent distributivity, compatibility and RRE, QOE. 
We show that, for the fixed state, distributivity of logic and compatibility are equivalent and that they are equivalent 
to RRE.  The result of paper \cite{PLOS} is made much stronger, even for the fixed mental state, QOE and RRE are incompatible 
(for the von Neumann-L\"uders measurements). We think that such state dependent treatment of observables is the right way to proceed in cognitive, psychological, and financial \cite{HavP} modeling. 

Since our paper is rather technical (although the proofs are placed in Appendixes A-B), it may be difficult to proceed through its first part. Paper's main output can be found in section \ref{RRET}. In particular, this section contains the description of an experimental test 
to check incompatibility of mental observables or in other words classicality (distributivity) of logic used in the process of decision making.

\section{Quantum reasoning}
\label{QR}

\subsection{Basics of quantum logic}
Operations of quantum logic are defined on the set $\cC(\cH)$ of closed subspaces of  a Hilbert space $\cH$, 
or equivalently on the set $\cP(\cH)$ of projections on  $\cH$.
\footnote{Hereafter, we call a closed subspace as a "subspace" for
brevity where no confusion may occur.
We do not specify the dimension of the underlying Hilbert space $\cH$.  
To understand this paper without any knowledge of mathematics for infinite dimensional Hilbert spaces,
the reader may assume that $\cH$ is finite dimensional.}
Subspaces (projections) are interpreted as mathematical representatives of propositions (events)
on a system under consideration. 

Let $P$ be a projection. Denote by $\cR(P)$ its range, i.e., $\cR(P)= P(\cH)$.  For a subspace $L$, 
denote by $\cP(L)$ the corresponding projection.  For a projection $P$, denote the projection onto the orthogonal complement of the subspace $\cR(P)$ by the symbol $\Not{P}$, i.e., $\cH=\cR(P) \oplus \cR(\Not{P})$
and $P\p=I-P$, where $I$ denotes the identity operator.  Negation of proposition $P$ is represented 
by $\Not{P}$.  The operations of  conjunction $\And$ and  disjunction $\Or$ are defined as follows.   

 Let $P$  and $Q$ be projections representing some propositions. The conjunction $P \And Q$ is defined as the projection on  the intersection of subspaces  $\cR(P)$ and $\cR(Q)$, i.e., $\cR(P \And Q) =  \cR(P) \cap \cR(Q)$. 
We remark that this operation is well defined even for noncommuting projections, i.e., incompatible quantum observables with values in $\{0,1\}$. Moreover, it is commutative:
\deq{
\label{EQ1}
P \And Q = Q \And P  
}
The same can be said about the operation of disjunction. Here the subspace $\cR(P\Or Q)$ is defined as the subspace generated by the union of subspaces $\cR(P)$ and $\cR(Q)$, i.e.,    $P\Or Q$ is the projection 
onto  this subspace. This operation is also well defined for non-commuting projections and, moreover, it is commutative:
\deq{
\label{EQ2}
P \Or Q = Q \Or P  
}
Thus logical operations of quantum logic is commutative. Typically this fact is not highlighted. Thus, in quantum reasoning noncommutativity is not present at the level of the basic operations of quantum logic, conjunction and disjunction.  

\section{Interrelation of distributivity and commutativity}
\label{SSS} 

\subsection{Two propositions}

We start with the simplest form of  distributive law, for  propositions $P$ and $Q$ and $R=\Not{Q}$,  
the negation of $Q$. We remark that, for any proposition $Q$, we have $Q\Or \Not{Q}= I$.
Then, the distributive law can be written in the form: 
\deq{
\label{EQ2a}
P= (P\And Q) \Or (P \And \Not{Q}),   
}
i.e.,
\deq{
\label{EQ2b}
P\And (Q \Or R)= (P\And Q) \Or (P \And R)),
}
in the same way
\deq{
\label{EQ2a1}
Q = (Q\And P) \Or (Q \And \Not{P}),   
}

The following theorem  \cite{BvN36} is a cornerstone of our modeling:

\medskip

{\bf Theorem A.}  The distributive law in the form (\ref{EQ2a}), (\ref{EQ2a1})
or equivalently 
\deq{
\label{EQ3a}
I = (P\And Q) \Or (P \And \Not{Q}) \Or (\Not{P} \And Q) \Or (\Not{P} \And \Not{Q})   
}
holds if and only if the projections $P,Q$ (multiplicatively) commute, i.e., $[P,Q]=0$, where 
$[P,Q]=PQ-QP$. 
\newcommand{\EQ}[1]{(\ref{#1})}

The above theorem was suggested by the observation of Birkhoff and von Neumann \cite[p.~833]{BvN36}
that $[P,Q]=0$ if and only if \Eq{EQ2a} holds.
A proof for the equivalence of
\Eq{EQ3a} and $[P,Q]=0$ will be given in Appendix \ref{se:Proof_A} 
  for the reader's convenience.

Thus, the distributivity for 4 propositions $P, Q,\Not{P},\Not{Q}$ is equivalent
to the commutativity for 2 projections $P,Q$.  
The operator defined by
\deq{
\label{EQ3bc}
\com(P,Q) = (P\And Q) \Or (P \And \Not{Q}) \Or (\Not{P} \And Q) \Or (\Not{P} \And \Not{Q})
}
is called the (quantum logical) commutator of $P$ and $Q$.
We set
\deq{
\label{EQ3b}
 d(P,Q) = I - \com(P,Q),
}
Note that Marsden \cite{Mar70} originally introduced $\d(P,Q)$ 
as the commutator of $P,Q$.
In quantum logic, quantities  $\com(P,Q)$ and  $d(P,Q)$ are the measures of distributivity and nondistributivity, 
respectively for propositions $P,Q,\Not{P}$, and $\Not{Q}$;  $\com(P,Q)=1$ or  $d(P,Q)=0$ in the distributive case.  According to Theorem A, the distributivity of propositions is equivalent to their compatibility, i.e., the possibility to assign their joint eigenvalues to them, say $(P,Q)=(0,0), (0,1), (1,0)$, or $(1,1)$.
In fact, the projections $P\And Q$, $P \And \Not{Q}$,
$\Not{P} \And Q$, and $\Not{P} \And \Not{Q}$
correspond, respectively,  to the joint eigensubspaces for $(P,Q)=(1,1)$,
 $(P,Q)=(1,0)$,  $(P,Q)=(0,1)$,  and $(P,Q)=(0,0)$. 
Thus, \Eq{EQ3a} means that every vector is a superposition of joint eigenvectors.

We remark that the operator $\com(P,Q)$ is Hermitian. By axioms for quantum theory it represents an observable. In particular, it is a projection, so that it is $\{0,1\}$-valued, or yes-no, observable. 
Thus, by Theorem A the distributive law can be represented via a quantum 
mechanical yes-no observable. Theoretically by measurement of this observable
it is possible to check the distributivity for two propositions. However, it seems to be difficult to present 
the  concrete measurement procedure of this observable. It reflects the similar problem with experimental checking of incompatibility.
Consider the Hermitian operator $i [P,Q]$. Theoretically by its measurement it is possible to check compatibility. However, the measurement procedure is not straightforward. 

Typically the lattice $\cP(\cH)$ of projections  is considered as a union of Boolean algebras, representing classical sub-logics of quantum logic. In this construction, the essence of Booleanity is commutativity of projections. 
Thus, commutativity of projections ensures that every inference rule in classical logic holds for all the sentences constructed by those commuting projections.
And distributivity of the basic operations, conjunction and disjunction, is the basic 
law of (classical) logic. Now, in view of Theorem A, we can characterize classical logic 
as the domains of validity of the distributive law. 

\subsection{Three propositions and their negations}

Consider now interrelation of  commutativity and distributivity for three propositions 
$\{P,Q,R\}$. Per definition the triple is commutative  if and only if each pair  $\{P,Q\}, \{P,R\}, \{Q,P\}$
is commutative. To couple commutativity and distributivity, we need to consider  not only 
these propositions, but  also their negations  $\Not{P},\Not{Q},\Not{R}$ (otherwise
the relation between commutativity and distributivity is not clear).  By distributivity of  $\{P,Q,R\}$ we mean the validity of  equality 
\deq{
\label{EQ2c}
X\And (Y \Or Z)= (X\And Y) \Or (X \And Z)), 
}
where $X,Y,Z=P,Q, R, \Not{P}, \Not{Q}, \Not{R}$. 

Triple $\{P,Q,R\}$ is commutative if and only (\ref{EQ2c}) holds, i.e., 
the lattice $\cP (P,Q,R)$ generated by $\{P, \Not{P},Q,\Not{Q},R,\Not{R}\}$ is distributive, or the ortholattice generated by $\{P,Q,R\}$ is a Boolean algebra. 
According to Bruns and Kalmbach \cite{BK73}, these conditions are equivalent to the equality
\deq{
\label{EQ2d}
\com(P,Q,R)=I,
}
where the (quantum logical) commutator $\com(P,Q,R)$ of $P,Q,R$ is given by
\begin{equation}
\label{EQ3T}
\com(P,Q,R)=(P \And Q \And R) \Or (P \And Q \And \Not{R}) \Or  (P \And \Not{Q} \And R) \Or  (P \And \Not{Q} \And \Not{R}) \Or 
(\Not{P}  \And Q \And R)
\end{equation}
$$
 \Or  (\Not{P}  \And Q \And \Not{R}) \Or  (\Not{P}  \And \Not{Q} \And R) \Or  (\Not{P}  \And \Not{Q} \And \Not{R}).
$$
We also introduce the measure of nondistributivity.
\deq{
\label{EQ3TT}
d(P,Q,R)= I - \com(P,Q,R)
}
It equals zero in the distributive case.

As in the case of  two propositions, the operator $\com(P,Q,R)$ is a projection
 and it represents a $\{0,1\}$-valued quantum observable. 
 But, we repeat that design of the corresponding measurement procedure is not straightforward.

\section{State dependent quantum logic}

\subsection{Quantum Platonism}

Reasoning based on the Birkhoff-von Neumann quantum logic (BvN-logic)  
is state-independent. 
It reflects intrinsic logic of interrelation between propositions based on the relation
$P\le Q$ ($P$ implies $Q$ ) if and only if  $\{P\}\subseteq\{Q\}$, where $\{P\}$
and $Q$ stand for the sets of states for which the proposition $P$ and $Q$ hold, 
respectively,  with certainty  \cite[p.~827]{BvN36}.
We can compare such viewpoint on propositions with Platonism as universals 
existing independently of particulars, in our case systems' states. The state-independent reasoning is an important 
area of information processing by humans, processing independent of human believes.  

\subsection{Distributivity}

\sloppy
Now, let us couple BvN-calculus (``quantum Platonic calculus'') to the states of mind - mental states. 
Then, for some states, the distributive law holds true 
\deq{
\label{EQ4}
P\psi= [(P\And Q) \Or (P \And \Not{Q})]\psi \quad\mbox{and}\quad
P\p\psi=[P\p\And Q)\Or(P\p\And Q\p)] \psi,  
}
or 
\deq{
\label{EQ5}
\psi = \com(P,Q) \psi, \quad \mbox{or equivalently} \quad d(P,Q) \psi =0, 
}
even if ``Platonic equalities'' of section \ref{SSS} are violated;
for three statements,
\deq{
\label{EQ5a}
\psi = \com(P,Q,R)\psi \quad \mbox{or equivalently} \quad d(P,Q,R) \psi =0.
}
Note that the above condition \eq{EQ5} or \eq{EQ5a} is equivalent to 
the condition that the measurement of the yes-no observable
$\com(P,Q)$ or $\com(P,Q,R)$, respectively, 
in the state $\psi$ always (with probability 1) leads to the ``yes'' result.
Consider the ortholattice $\cP (P,Q,R)$ generated by
$P,Q,R$. 
If condition (\ref{EQ5a}) holds, then the lattice $\cP (P,Q,R)$ is distributive for the state $\psi:$
\deq{
\label{EQ5b}
X\And (Y \Or Z) \psi= [(X\And Y) \Or (X \And Z)]\psi, 
}
where $X,Y,Z=P,Q, R, \Not{P}, \Not{Q}, \Not{R}$. 
We set
\deq{
L_{Q,P,R} =\{\psi \in \cH: d(P,Q, R) \psi=0\},
}
the kernel of the operator $d(P,Q, R)$. 
This is a linear subspace of $\cH$. 
\sloppy
Moreover, $L_{P,Q, R}$ is a common invariant subspace of $P,Q, R$, and hence by
replacing $P,Q, R$ by $PL_{P,Q, R}, QL_{P,Q, R}, RL_{P,Q, R}$, projections $P,Q, R$
act on $L_{P,Q, R}$ as mutually commuting projections. 
We call it {\it the distributivity subspace} of the lattice  
$\cP (P,Q,R)$. For states from this subspace,   logic of reasoning
is classical. We remark that such classicality is the  delicate 
issue. Logic of propositions in $\cP (P,Q,R)$ can be nonclassical, i.e., 
$
d(P,Q,R)= I - \com(P,Q,R)
$
can be nonzero. But, for states from $L_{Q,P,R}$, reasoning is classical -- the 
distributive law holds true. 
 
This is good place to remark that 
\deq{
L_{Q,P,R}= \com(P,Q,R) \cH.
}
In fact, the lattice $\cP (P,Q,R)$ is acting on $L_{Q,P,R}$ as a Boolean algebra,
or  the lattice $\cP (PL_{P,Q, R}, QL_{P,Q, R}, RL_{P,Q, R})$ is indeed
a Boolean algebra. 

\subsection{General theory}
The (quantum logical) commutator of two projections $P,Q$  was 
originally introduced by Marsden \cite{Mar70} by the dual form to \Eq{EQ3bc} as 
 \deq{
 d(P,Q)=(P\Or Q) \And (P \Or Q\p) \And (P\p \Or Q) \And (P\p \Or Q\p),
 }
 which is equivalent to \Eq{EQ3b}.
 We follow the recent convention to call $\com(P,Q)$ the (quantum logical) 
 commutator of two projections $P,Q$.
 Bruns and Kalmbach \cite{BK73} extended this notion to any finite set $\cF$ of 
 propositions by
 \deq{\label{BK73}
\com(\cF)=\Sup_{\al:\cF\to\{\id,\perp\}}\Inf_{P\in\cF}P^{\al(P)},
}
where 
$\{\id,\perp\}$ stands for the set consisting of the identity operation $\id$ and the 
orthocomplementation~$\perp$.
Note that for $\cF=\{P,Q,R\}$,   \Eq{BK73} reduces to \Eq{EQ3T}.
Generalizing this notion to arbitrary sets $\cA$ of propositions,
Takeuti \cite{Ta81} defined the commutator $\com(\cA)$ of $\cA$  by
\deq{\label{Ta81}
\com(\cA)=\Sup \{E\in\cP(\cH) \mid \mbox{$[E,P]=0$ and $[P,Q]E=0$ 
for all $P,Q\in\cA$}\}.
}
Subsequently, Pulmannov\'{a} \cite{Pul85} proved 
the relation
\deq{
\com(\cA)=\Inf\{\com(\cF)\mid \mbox{$\cF$ is a finite subset of $\cA$}\}.
}
Thus, $\com(\cA)$ is the limit of $\com(\cF)$ for $\cF\to \cA$.

From \Eq{Ta81}, it can be seen that the subspace $L_{\cA}=\com(\cA)\cH$ is the 
maximum $\cA$-invariant subspace of $\cH$ such that any pair $P,Q$ in $\cA$ 
commute on $L_{\cA}$.  

The notion of commutators and the state-dependent perspective of quantum
logic play a crucial role in the recent study of quantum set theory 
\cite{Ta81,07TPQ,16A2,17A2,21QTD}, which reconstructs quantum
theory in the mathematical universe based on quantum logic with a close
connection to the topos approach to quantum theory \cite{21BBQ}.

\section{Value function}

As in any logical reasoning, it is useful to explore a {\it value function}. In 
quantum logic, it is based on quantum probability defined 
by the Born's rule
\deq{
\label{EQ6}
\rm{Pr}\{P\| \psi\} = \Vert P \psi \Vert^2.  
}
This value function is explored in the  process of reasoning. 
The proposition $P$ holds in the state $\psi$ 
iff  $\Pr\{P\| \psi\}=1$. In particular, the distributive law holds 
in $\psi$ if 
\deq{
\label{EQ6X}
\Pr\{\com(P,Q,R) \| \psi\} = \Vert \com(P,Q,R) \psi \Vert^2 =1.  
}

Now let us characterize (equivalent) conditions for validity of the statement: ``conjunction $P\And Q$ of the propositions $P$ and $Q$ holds in the state $\psi$'',

\medskip
{\bf Theorem B.} The following conditions are all equivalent.
\begin{enumerate} 
\item $\rm{Pr}\{P\And Q \| \psi\}=1;$
\item $P$ holds in $\psi$ and simultaneously $Q$ holds in $\psi;$
\item $\rm{Pr}\{P\|\psi\}=\rm{Pr}\{Q\|\psi\}=1;$ 
\item (i) $P$ and $Q$ are commuting in $\psi$, i.e., $[P,Q]\psi=0$,
and (ii) $P$ holds in $\psi$ and simultaneously $Q$ holds in $\psi$, i.e.,  
\deq{
\label{EQ88}
\Pr\{\com(P,Q)\| \psi\}=\Pr\{P\|\psi\}=\Pr\{Q\|\psi\}=1.
}
\end{enumerate}

We shall give a proof  in Appendix \ref{se:Proof_B} for the reader's convenience, although the assertion is rather well-known \cite{21QTD}.

It has often been claimed that the state-independent interpretation of 
the conjunction $P\And Q$ in quantum logic is ambiguous if projections 
$A$ and $B$ do not commute.
However, Theorem B shows that the state-dependent interpretation as 
 ``$P \And Q$ holds in the state $\psi$" is unambiguous, since 
if ``$P \And Q$ holds in the state $\psi$" then $P$ and $Q$ are commuting
in the state $\psi$, i.e., two projections $P$ and $Q$ are actually commuting
on the relevant subspace $L_{P,Q}$. 
For further discussions on the state-dependent interpretations of other
logical operations in quantum logic, we refer the reader to \cite[Section 5]{21QTD}. 

\section{RRE as experimental test of distributivity of human logic}
\label{RRET}

\subsection{Notion of response replicability}

We recall that the projective instrument $\cI_P$
is a family $\{\cI_P(x)| x=0,1\}$ of positive maps $\cI_P(x)$ on the space
$L(\cH)$ of operators on $\cH$ defined by the relation $\cI_P(1)\rho=P\rho P$ and $\cI_P(0)\rho=P\p \rho P\p$ for all $\rho\in L(\cH)$.  See \cite{ENTROPY,MATHPSYCH} for further information.

We consider two projections $P,Q$, and their projective instruments $\cI_P$ and
$\cI_Q$, and a state vector $\psi$. Denote their output probability by 
\deq{
p(Xx,Yy,Zz,..)&=\Tr(\cdots \cI_Z(z)\cI_Y(y)\cI_X(x)\ketbra{\psi})
}
for $X,Y,Z\in\{P,Q\}$ and $x,y,z\in\{0,1\}$.
Then we have
\deq{\label{eq:PM}
p(Xx,Yy,Zz,..)&=\|\cdots Z^{(z)}Y^{(y)}X^{(x)}\psi\|^2,
}
where $X^{(0)}=X\p$ and $X^{(1)}=X$ for all $X\in\{P,Q\}$.
The instruments $\cI_P$ and $\cI_Q$ show the  repeatability, i.e., $p(Px,Px)=p(Px)$
and $p(Qx,Qx)=p(Qx)$ for any $P,Q$ and $\psi$.
We say that  $\cI_P$ and $\cI_Q$ show {\em RRE (the response replicability effect) 
in  $\psi$} iff $p(Px,Qy,Px)=p(Px,Qy)$ and $p(Qx,Py,Qx)=p(Qx,Py)$ for any
$x,y,z\in\{0,1\}$, i.e., 
\deq{
 p(P1,Q1,P1)&=p(P1,Q1),\label{eq:1}\\  
 p(P1,Q0,P1)&=p(P1,Q0),\\ 
 p(P0,Q1,P0)&=p(P0,Q1),\\ 
 p(P0,Q0,P0)&=p(P0,Q0),\label{eq:4}\\ 
 p(Q1,P1,Q1)&=p(Q1,P1),\label{eq:5}\\ 
 p(Q1,P0,Q1)&=p(Q1,P0),\\
 p(Q0,P1,Q0)&=p(Q0,P1),\\ 
 p(Q0,P0,Q0)&=p(Q0,P0).\label{eq:8}
}

\subsection{Equivalence of distributivity and RRE}

The following theorem holds: 

\bTheorem\label{th:EDR}
The projective instruments of $P$ and $Q$ show RRE  in  a state $\psi$ if and only if
$\com(P,Q)\psi=\psi$.
\eTheorem
The proof will be given in Appendix \ref{se:Proof=EDR}. 

Since the above proof of  ($\THEN$) uses only Eqs.~\eq{eq:1}--\eq{eq:4}, we have
\bCorollary
In the definition of RRE,
Eqs.~\eq{eq:1}--\eq{eq:4} imply Eqs.~\eq{eq:5}--\eq{eq:8}.
\eCorollary

From the above theorem, we can test the distributivity of human logic (or by Theorem A, commutativity of projections $P$ and $Q)$
in a given state $\psi$ by two projective instruments $\cI_P$ and $\cI_Q$; namely,
$P$ and $Q$ commute in $\psi$ if and only if 
$\cI_P$ and $\cI_Q$ show RRE, 
or equivalently $\cI_P$ and $\cI_Q$ show Eqs.~\eq{eq:1}--\eq{eq:4}.

\subsection{Towards testing distributivity of human logic}

In this paper, we highlight the role of the distributive law in human reasoning. We couple violations of classical logic with violation of distributivity. Theorem \ref{th:EDR} provides (really unexpected) possibility to dive into the deepest level of human information processing. RRE can be checked experimentally, see  Eqs. (\ref{eq:1})-(\ref{eq:8}) (in fact, it is sufficient to check Eqs.~\eq{eq:1}--\eq{eq:4}). We hope that coupling of RRE with logic of human reasoning will stimulate psychologists to perform new experiments. In the light of paper \cite{WB13}, analysis of the methodology and design should precede experiment. By finding experimental violation of one of  Eqs.~\eq{eq:1}--\eq{eq:8}, experimenters can conclude that 
\begin{itemize}
\item either the distributive law is violated (in the state $\psi$ prepared for the experiment),
\item or the state update generated by observations cannot be described straightforwardly as orthogonal projection.
\end{itemize} 

We stress that {\it the projective-state update implies the classical Bayesian update of probability}
for two commuting observables \cite{Hav16}
and the use of Bayesian inference in reasoning.  Non-projective  instruments generate the non-Bayesian state updates and new inference procedures \cite{ENTROPY,MATHPSYCH}. 

\subsection{Towards testing  incompatibility of mental observables}

As was already pointed out, it is difficult if possible at all to prove compatibility (incompatibility) of mental observables 
in the theoretical framework (see introduction). It seems that it can be determined only experimentally. Since commutativity is equivalent to distributivity, the RRE-test can be used as well for checking  compatibility (incompatibility) of projection observables. 

\subsection{Impossibility of description by projective instruments of combination QOE+RRE for one concrete state}

All the previous considerations of this section were done in the state-dependent framework. This gives the possibility to improve 
essentially the basic result \cite{PLOS} on the impossibility to describe  combination of QOE and RRE by projective type instruments. This no-go theorem was formulated under the following {\it stability assumption}:

``If  $\psi$ is a possible initial state vector for a given measurement sequence in an $n$-dimensional Hilbert space, then there is an open ball $B_r(\psi)$ centered at $\psi$  with a sufficiently small radius $r>0$, such that any vector $\psi + \delta$ in this ball, normalized by its length $\Vert \psi + \delta\Vert$, is also a possible initial state vector for this measurement sequence.''

Now, we can omit this stability condition and consider just one fixed state $\psi$. If QOE+RRE holds for this state, then measurements cannot be described by projective instruments. 

\section{Concluding remarks}

We discuss the conjecture that quantum logic is a tool of human reasoning; the brain functioning includes the special system for 
information processing based on quantum logic, the QL-system. The role of violation of distributivity is highlighted.
As is shown (Theorem A),  distributivity is equivalent to compatibility (and in the quantum formalism, to 
commutativity of operators). 

The state-dependent character of quantum reasoning is emphasized. We think that state-dependent modeling of quantum reasoning is especially important for applications to cognition and psychology. 

Our present study is closely coupled to the previous research on quantum-like modeling of QOE and RRE \cite{ENTROPY,MATHPSYCH}. 
The role of RRE was highlighted through coupling to quantum reasoning, its (non-)distributivity and using (in)compatible logical variables. We proposed 
the experimental test for RRE and it can be considered as a test of distributivity-compatibility under the assumption of projective type representation of mental observables. Finally, we improved the no-go theorem from article \cite{PLOS}, on the impossibility of combination of QOE and RRE.

It is often questioned whether POVMs (positive [or, probability] operator valued measures) are relevant to the present problem.
First of all, we should note that it is sometimes claimed that all measurements are classified 
as projective measurements or POVM measurements.  
However, this is not a correct classification of measurements, since the notion of projective measurements 
implies that the state is updated by the measurement by the projection onto the eigensubspace determined by
the measurement outcome and measured observable, but the notion of POVM does not imply 
any particular way of the state update.
Thus, it is appropriate to say that 
all measurements are classified as sharp measurements or unsharp measurements. 
Every sharp measurement has a projection valued measure 
to determine the probability distribution of the outcome of the measurement
and every unsharp measurement has a POVM that is not a projection valued measure.  
Now, sharp measurements are classified as repeatable sharp measurements 
and non-repeatable sharp measurements, where ``repeatability'' is synonymous with ``$A-A$ response replicability''.
All repeatable sharp measurements are classified as projective (or non-invasive repeatable sharp) 
measurements or invasive repeatable sharp measurements.  In our previous
paper we have shown that a pair of invasive repeatable sharp measurements
shows RRE and QOE consistently.  Thus, invasive repeatable sharp
measurements are the most appropriate class to model opinion poll data.
This class of measurements are well-described as quantum instruments, which are derived
from general quantum measurement theory and do not satisfy the Bayesian
belief-update rule \cite{Oza84,Oza21}.  
Unsharp positive operator valued measures are not relevant, 
since they do not show repeatability in the finite dimensional case \cite{BDP04} (see also \cite[Theorem 6.5]{Oza84}).

In this direction of research the following open problem remains:

{\em Are there any quantum instrumental model $(\cI_P, \cI_Q)$ of QOE+RRE 
such that $P$ and $Q$ do not commute.}

The Wang-Busemeyer model \cite{WB13} satisfies QOE,  and is such that 
$P$ and $Q$ do not commute, but violates RRE, 
as their model is of the projective update type.  
In this paper, we have concluded, in any models $(\cI_P, \cI_Q)$ of the projective update type, QOE implies the noncommutativity of $P$ and $Q$, but  RRE implies the
commutativity of $P$ and $Q$. 
Our model \cite{ENTROPY,MATHPSYCH} satisfies  both QOE and RRE,
but $P$ and $Q$ in the model commute.  It is an interesting problem as to whether 
there is some type of opinion poll data sets that demands models
$(\cI_P, \cI_Q)$  of  QOE+RRE such that $P$ and $Q$ do not commute.

We hope that this paper would attract attention of psychologists and experts in brain studies to  quantum logic conjecture for human reasoning. In experimental research, coupling of RRE with the basics of quantum reasoning would stimulate its further testing 
(cf. \cite{BWRRE}).   

The present research, though qualitative, motivates a further quantitative
research. We expect that the tests of the RRE involves measurement error,
but the deviation from RRE by this error is too small to explain the
degree of QOE that manifests in the existing poll data.  Thus, projective
measurement model would not explain RRE+QOE (in the poll data) even
compromised by the error.

\section*{Acknowledgments}
This work was partially supported by  JSPS KAKENHI Grant Numbers JP22K03424,
JP21K11764, JP19H04066.

\appendix
\section{Proof of Theorem A}\label{se:Proof_A}
\bProof
Since  $P\And Q$, $P \And \Not{Q}$, $\Not{P} \And Q$, and $\Not{P} \And \Not{Q}$ are mutually orthogonal, we have
$$
(P\And Q) \Or (P \And \Not{Q}) \Or (\Not{P} \And Q) \Or (\Not{P} \And \Not{Q})= (P\And Q) + (P \And \Not{Q}) + (\Not{P} \And Q) + (\Not{P} \And \Not{Q}).   
$$

($\Then$)  
Since $\cR(P\And Q)\subseteq\cR(P)$, we have $P(P\And Q)=P\And Q$.  Similarly, 
$Q(P\And Q)=P\And Q$. Thus, $PQ(P\And Q)=QP(P\And Q)=P\And Q$.  Similarly, we have
 $PQ(P\And Q\p)=QP(P\And Q\p)=0$, $PQ(P\p\And Q)=QP(P\p\And Q)=0$, and
  $PQ(P\p\And Q\p)=QP(P\p\And Q\p)=0$.
  Thus, the assertion follows easily.
  
($\If$) It is well known that if $[P,Q]=0$ then $P\And Q=PQ$, $P\p\And Q=(I-P)Q=Q-PQ$,
$P\And Q\p=P(I-Q)=P-PQ$, $P\p\And Q\p=(I-P)(I-Q)=I-P-Q+PQ$.  Thus, the assertion follows
easily.
\eProof

\section{Proof of Theorem B}\label{se:Proof_B}
\bProof
$(1)\Then(4)$:  From (1) we have $\|(P\And Q)\psi\|^2=1$.
Since $ P\And Q \le  \com(P,Q)$, we have $1= \|(P\And Q)\psi\|^2\le\|\com(P,Q)\psi\|^2$,
so that $\Pr\{\com(P,Q)\|\psi\}=1$.  Similarly, $\rm{Pr}\{P\|\psi\}=1$ follows from
$ P\And Q \le P$, and  $\rm{Pr}\{Q\|\psi\}=1$ from $ P\And Q \le Q$. Thus, the 
implication $(1)\Then(4)$ follows.

Now, the implications $(4)\Then(3)$, $(3)\Then(2)$, and $(2)\Then(1)$ can be
shown easily.
\eProof

\section{Proof of Theorem \ref{th:EDR}}\label{se:Proof=EDR}
\bProof
($\If$) 
Suppose $\com(P,Q)\psi=\psi$.  Then 
\deqs{
QP\psi&=QP\com(P,Q)\psi=Q[(P\And Q)\Or(P\And Q\p)]\psi=(P\And Q)\psi,\\
PQP\psi&=P(P\And Q)\psi=(P\And Q)\psi.
}
Thus, we have $PQP\psi=QP\psi$. Similarly, 
we have $X^{(x)}Y^{(y)}X^{(x)}\psi=Y^{(y)}X^{(x)}\psi$,
and RRE follows.

 ($\THEN$) From \Eq{eq:1}, $\|PQP\psi\|^2=\|QP\psi\|^2$.
 Since $\|QP\psi\|^2=\|PQP\psi\|^2+\|P\p QP\psi\|^2$, we have $P\p QP\psi=0$,
 and hence $QP\psi=PQP\psi$.  Thus, $QP\psi\in\cR(P)\cap\cR(Q)$,
 and hence $(P\And Q) QP\psi=QP\psi$,
so that $\com(P,Q)QP\psi=QP\psi$.  Similarly, from Eqs.~\eq{eq:1}--\eq{eq:4} we have 
$\com(P,Q)Q^{(y)}P^{(x)}\psi=Q^{(y)}P^{(x)}\psi$. Thus, we have
\deqs{
\com(P,Q)\psi&=\com(P,Q)\sum_{x,y}Q^{(y)}P^{(x)}\psi
=\sum_{x,y}\com(P,Q)Q^{(y)}P^{(x)}\psi\\
&=\sum_{x,y}Q^{(y)}P^{(x)}\psi=\psi.}
Therefore, we conclude that the projective measurements of 
$P$ and $Q$ show RRE in $\psi$ if and only if $\com(P,Q)\psi=\psi$.
\eProof

\end{document}